\documentclass[hidelinks,onefignum,onetabnum]{siamart250211}

\usepackage{enumitem}

\usepackage{lipsum}
\usepackage{amsfonts}
\usepackage{graphicx}
\usepackage{epstopdf}
\usepackage{algorithmic}
\ifpdf
  \DeclareGraphicsExtensions{.eps,.pdf,.png,.jpg}
\else
  \DeclareGraphicsExtensions{.eps}
\fi

\newsiamremark{remark}{Remark}
\newsiamremark{hypothesis}{Hypothesis}
\crefname{hypothesis}{Hypothesis}{Hypotheses}
\newsiamthm{claim}{Claim}
\newsiamremark{fact}{Fact}
\crefname{fact}{Fact}{Facts}

\headers{Benchmarking 1D QLBM on Quantum Hardware}{M. Quartey and L. Zhang}

\title{Benchmarking One-Dimensional Quantum Lattice Boltzmann Methods on Quantum Hardware\thanks{Submitted to the editors July 2026.
\funding{This work was funded by IBM under Artificial Intelligence Research Collaboration Agreement contract no.~W1771793.}}}

\author{Melvina Quartey\thanks{Department of Mechanical, Aerospace, and Nuclear Engineering, Rensselaer Polytechnic Institute, Troy, NY 
  (\email{quartm2@rpi.edu}, \email{zhanglucy@rpi.edu}, \url{https://www.lucy-t-zhang.org/}).}
\and Lucy T. Zhang\footnotemark[3]}

\usepackage{amsopn}

\ifpdf
\hypersetup{
  pdftitle={Benchmarking One-Dimensional Quantum Lattice Boltzmann Methods on Quantum Hardware},
  pdfauthor={M. Quartey, L. Zhang}
}

\begin{document}

\maketitle

\tableofcontents
\begin{abstract}
Classical computational fluid dynamics (CFD) methods encounter significant computational bottlenecks in large systems and at high Reynolds number flows due to the rapid increase in degrees of freedom, even on flagship supercomputers. As quantum computers emerge, the Quantum Lattice Boltzmann Method (QLBM) has been proposed as a quantum-native alternative for solving fluid dynamics problems. However, until recently, nearly all studies have relied on classical quantum simulators rather than real quantum hardware. This study presents a benchmark of a one-dimensional QLBM implementation executed on three different IBM Quantum hardware. We design and test a quantum circuit for the advection-diffusion equation using amplitude encoding, a Bhatnagar–Gross–Krook (BGK)-type collision operator, and a streaming step with periodic boundary conditions. Our numerical experiments evaluate transpiled two-qubit gate count, total circuit depth, runtime, and accuracy metrics including maximum absolute error ($\mathcal{L}_\infty$ norm) and root-mean-square error (RMSE). We find that the present collision and streaming operators dominate the overall cost and become the primary scalability bottleneck on current hardware. To mitigate the prohibitive cost of full quantum state tomography on hardware-based benchmarking exercises, we introduce a sampler-based measurement strategy that reconstructs macroscopic density fields directly from computational-basis measurements. This approach substantially reduces measurement overhead while preserving the expected advection–diffusion dynamics on hardware. These results provide a quantitative assessment of QLBM performance in the noisy intermediate-scale quantum (NISQ) regime and identify the principal algorithmic and experimental limitations that must be addressed for QLBM to become viable for fluid dynamics.
\end{abstract}



\section{Introduction}
\label{sec:intro}

\subsection{Background}
Computational Fluid Dynamics (CFD) enables meaningful fluid simulations and studies in science and engineering. Various CFD models exist to solve specific cases of the Navier–Stokes equations to numerically predict the behaviors of complex fluid systems in many industries, including aerospace, automobile, and environmental engineering. As the domain size and behavioral complexity of flows increase, the number of degrees of freedom (DoFs) and subsequent number of floating point operations (FLOPS) required to resolve them scale rapidly~\cite{Pope_2015}. Presently, our ability to process large-scale high-fidelity simulations remains limited by the computational resources of our most advanced exascale classical supercomputers~\cite{Syamlal_Copen_Takahashi_Hall_2024}. 

As technology develops, researchers seek improved algorithms to solve complex multiscale fluid phenomena, such as turbulent flows, while optimizing computational costs. Total computational complexity of high-fidelity fluid turbulence simulation, such as the direct numerical simulation (DNS) model, scales with Reynolds number ($Re$) as $O(Re^3)$~\cite{Moin_Mahesh_1998}. Real-world applications in aerodynamics typically involve flows of $O(10^8)$ $Re$~\cite{Spalart_2000}. This is taxing for even modern high-performance computers (HPCs) as $O(Re^{9/4})$ spatial degrees of freedom are required to resolve down to the smallest eddies in such models~\cite{Kolmogorov_1991}. As a result, simulating high-fidelity flows within large-scale domains remains a grand challenge in CFD.

Quantum computing offers a fundamentally different computational paradigm. It exploits superposition and entanglement to encode and process information in exponentially large Hilbert spaces~\cite{Nielsen_Chuang_2010}. This presents a promising solution to these limitations, specifically through Quantum Computational Fluid Dynamics (QCFD) algorithms. Early work on quantum algorithms for partial differential equations (PDEs) includes the Quantum Lattice Gas Automata~\cite{Yepez_2001}, the Harrow–Hassidim–Lloyd algorithm (HHL) for linear systems~\cite{Harrow_Hassidim_Lloyd_2009}, quantum variational eigensolver (VQE) approaches to differential operators~\cite{Lubasch_Joo_Moinier_Kiffner_Jaksch_2020}, and Carleman‐linearization methods for the Navier–Stokes equations~\cite{Liu_Kolden_Krovi_Loureiro_Trivisa_Childs_2021}. 

In recent years, authors have investigated hybrid quantum-classical algorithms~\cite{Kyriienko_Paine_Elfving_2021, Malinverno_2023}, Quantum-assisted Physics-Informed Neural Networks (PINNs), and the Quantum Lattice Boltzmann Method (QLBM)~\cite{Budinski_2021}. Among quantum‐native CFD approaches, the Lattice Boltzmann Method (LBM) stands out because of its mesoscopic, particle distribution representation, which naturally maps to quantum amplitudes and unitary operations~\cite{Wawrzyniak_Winter_Schmidt_Indinger_Janssen_Schramm_Adams_2025}. The classical LBM updates distribution functions on a discrete velocity set via collision, streaming, and macroscopic reconstruction steps~\cite{Mohamad_2019}; this structure maps favorably to quantum circuits, helping to preserve locality and structured update rules on a gate-based quantum computer. The LBM here offers a kinetic theory reformulation of the Navier-Stokes equations governing fluid mechanics~\cite{Kruger_Kusumaatmaja_Kuzmin_Shardt_Silva_Viggen_2017}. Recent work has shown that QLBM can achieve logarithmic scaling in qubit number and polynomial scaling in gate count for certain flow models, indicating a potential route toward quantum advantage in fluid dynamics~\cite{Syamlal_Copen_Takahashi_Hall_2024}.

Despite the promise of QLBM towards reliable QCFD, several bottlenecks are encountered on the analysis of a typical algorithm. Hardware implementation of QCFD algorithms is a major challenge, as modern quantum computers are characterized as Noisy Intermediate-Scale Quantum (NISQ) devices due to their limited error-handling capabilities. In QLBM, examples of these bottlenecks include the polynomial scaling of total circuit depth, the exponential number of circuit evaluations performed during measurement via full state tomography, the rapid loss of state fidelity as the simulation evolves, and the impractically long runtimes even for small domains. A massive research effort is being made to address these challenges on all fronts~\cite{Tiwari_Iaconis_Jojo_Ray_Roetteler_Hill_Pathak_2025, Wang_Meng_Zhao_Yang_2025, Shinde_Budinski_Niemimaki_Lahtinen_Liebelt_Li_2025, Lacatus_Moller_2025}.

\subsection{Roadmap}
In this work, we address these challenges by implementing a QLBM algorithm on three different IBM quantum processors using the Qiskit 2.3 software development kit. We focus on a one-dimensional QLBM formulation of the advection–diffusion equation, which serves as a minimal but physically meaningful model of transport and spreading processes.

Our implementation follows the standard QLBM workflow: quantum state preparation of particle distribution functions, a Bhatnagar–Gross–Krook (BGK)-type collision operator implemented as a unitary transformation, quantum streaming with periodic boundary conditions, and measurement to extract macroscopic density fields. By executing the full algorithm on real quantum hardware, we provide a hardware-level assessment of circuit depth, gate counts, runtime, and numerical accuracy. We further introduce a sampler-based measurement strategy that reduces the prohibitive overhead associated with full quantum state tomography. This work bridges the gap between an abstract QLBM algorithm design and the practical constraints imposed by NISQ-era quantum processors.

The remainder of this paper is organized as follows: \cref{sec:lbm_method} reviews the LBM method and its formulation for the one-dimensional advection–diffusion equation. Section 3 presents the QLBM algorithm and analyzes its computational complexity, with a particular focus on the measurement step. Section 4 describes the experimental setup, including circuit construction, quantum backends, and performance metrics. Section 5 reports and analyzes numerical results obtained on hardware backends and compares them to the IBM Qiskit simulator, highlighting scaling behavior and hardware limitations. Finally, Section 6 summarizes the main findings and outlines directions for future research.

\section{Preliminaries}
\label{sec:prelims}

\subsection{The Lattice Boltzmann Method}
\label{sec:lbm_method}
 Lattice Boltzmann Method (LBM) is a mesoscopic numerical framework that models transport phenomena through the evolution of particle distribution functions on a discrete lattice. Rather than solving macroscopic governing equations directly, LBM evolves populations of fictitious particles whose collective behavior recovers the desired continuum dynamics. A key to note about LBM is that it discretizes the Boltzmann equation in space, time, and velocity.

Let $f_i(\mathbf{x}, t)$ denote the particle distribution function associated with the lattice position $\mathbf{x}$, the time $t$, and the discrete velocity set $\mathbf{c}_i$. Here, $\mathbf{c}_i$ represents the discrete directions that particles in a lattice are transported along with some probability. The standard Lattice Boltzmann equation with a single-relaxation-time BGK collision operator is
\begin{equation}
f_i(\mathbf{x} + \mathbf{c}_i \Delta t, t + \Delta t) = f_i(\mathbf{x}, t) - \frac{1}{\tau} \left[ f_i(\mathbf{x}, t) - f_i^{eq}(\mathbf{x}, t) \right],
\label{eq:bgk}
\end{equation}
where $\tau$ is the relaxation time controlling dissipation and $f_i^{eq}$ is the local equilibrium distribution. 

In LBM, $D_aQ_b$ is a notation used to classify different lattice configurations based on their dimensionality and velocity directions. ``$a$'' defines the number of spatial dimensions resolved, and ``$b$'' defines the number of discrete velocity directions used to describe the transport phenomena. Some examples are D1Q3, D2Q9, and D3Q27. 

For transport problems such as advection–diffusion, the equilibrium distribution is typically chosen as a low-order expansion in the macroscopic velocity $u$. Here, for our D1Q3 lattice with discrete velocities $c_i$, we will use
\begin{equation}
f_i^{eq} = w_i \rho \left( 1 + \frac{c_i u}{c_s^2} \right),
\end{equation}
where $\rho$ is the density, $u$ is the prescribed advection velocity, $w_i$ are lattice weights satisfying $\sum w_i = 1$, and $c_s$ is the lattice speed of sound. Macroscopic moments are recovered from the particle distribution functions as
\begin{equation}
\rho = \sum_i f_i, \quad \rho u = \sum_i f_i c_i.
\end{equation}

\subsection{Advection-Diffusion}
\label{sec:adv_diff}
The physical model considered in this work is described by the one-dimensional advection–diffusion equation,
\begin{equation}
\frac{\partial \rho}{\partial t} + u \frac{\partial \rho}{\partial x} = D \frac{\partial^2 \rho}{\partial x^2},
\end{equation}
where $\rho$ is the density, $u$ is the advection velocity, and $D$ is the diffusion coefficient. Within the LBM framework, this macroscopic equation is recovered through a Chapman–Enskog expansion of the Lattice Boltzmann equation under diffusive scaling~\cite{Kruger_Kusumaatmaja_Kuzmin_Shardt_Silva_Viggen_2017}. The diffusion coefficient is related to the relaxation time by
\begin{equation}
D = c_s^2 \left( \tau - \frac{1}{2} \right) \Delta t,
\end{equation}
linking physical transport properties directly to numerical parameters.

\subsection{Problem Definition}
\label{sec:problem_def}
\begin{figure}[H]
    \centering
    \includegraphics[width=0.8\textwidth]{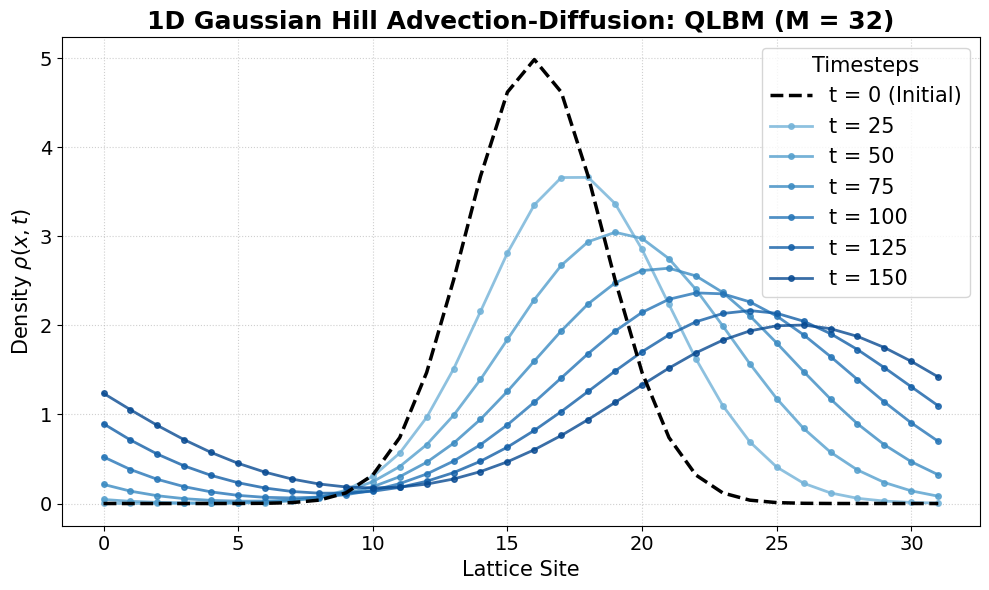}
    \caption{Ideal QLBM simulation of Gaussian hill advection-diffusion at $M=32$ over 150 timesteps.}
    \label{fig:gaussian_hill}
\end{figure}

In this work, we consider the one-dimensional advection–diffusion of a Gaussian initial concentration profile solved using a D1Q3 Lattice Boltzmann formulation. We simulate a physical domain of unit length, $L = 1.0$, discretized into a lattice with spacing
\begin{equation}
\Delta x = \frac{L}{M},
\end{equation}
where $M$ represents the number of lattice sites in the domain. The lattice model employs three discrete velocities, $c_i = \{ 0, 1, -1 \}$, corresponding to resting, right-moving, and left-moving particle populations, with lattice weights $w_0 = 2/3, w_1 = 1/6, w_2 = 1/6$. The lattice sound speed is given by $c_s^2 = 1/3$, and an initial uniform velocity is prescribed across the domain. The equilibrium distribution function for this D1Q3 model is defined as
\begin{equation}
    f_i^{eq} = w_i\rho  \left( 1 + \frac{c_i u}{c_s^2} \right).
\end{equation}
Macroscopic density is recovered from the particle populations via $\rho = \sum_i f_i$. The Gaussian initial density profile is given as
\begin{equation}
    \rho(x, 0) =  \exp \left( -\frac{(x-x_0)^2}{2\sigma^2} \right),
\end{equation}
centered at $x_0$ with a width parameter $\sigma = 0.08L$. Periodic boundary conditions are imposed at the domain boundaries. 
When solved ideally on a quantum simulator for 150 timesteps, the density profile evolution for the $M = 32$ case is illustrated in \cref{fig:gaussian_hill}.

At the end of each timestep, we extract the macroscopic density and velocity by 

\begin{equation}
    \rho = \sum_i f_i, \quad u = \frac{1}{\rho} \sum_i f_i c_i  \label{eq:moments}
\end{equation}
 and perform updates for the next timestep. We conduct numerical experiments on lattice sizes $M=4$ up to $M=64$ to assess resolution effects and hardware scaling behavior.

\section{Algorithm}
\label{sec:alg} 

\subsection{Quantum Lattice Boltzmann Method}
The QLBM maps the classical LBM update rule onto a quantum circuit that evolves a quantum state encoding $f_i$. For a D1Q3 lattice, the quantum state at time step $t$ is represented as:
\begin{equation}
    |\psi\rangle_t = \sum_{i=0}^{2}\sum_{j=0}^{M-1}  \sqrt{f_i(x_j, t)} |i\rangle_v|j\rangle_s ,
\end{equation}
where the $n_s = \log_2 M$ qubit register encodes the lattice positions in space, and the velocity register $n_v$ encodes the discrete velocity directions. We follow the general QLBM time step consisting of the following four stages~\cite{Kocherla_Song_Chrit_Gard_Dumitrescu_Alexeev_Bryngelson_2024}:

\begin{description}[
    leftmargin=3em,
    labelindent=3em,
    labelwidth=3em,
    align=left
]
    \item[i. Initialization] \hfill \\
    The initial macroscopic density, $\rho(x, 0)$, is expressed in terms of particle distribution function and encoded into the amplitudes of the quantum state such that

        \begin{equation}
        f_i(x,0) = w_i\rho(x,0),
    \end{equation}

    and
    
    \begin{equation}
        |\psi\rangle_0 = \sum_{i=0}^{2}\sum_{j=0}^{M-1} \sqrt{f_i(x_j, 0)} |i\rangle_v|j\rangle_s .
    \end{equation}
    Here, the qubits used are the spatial and velocity qubits, $s$ and $v$ respectively.

    \item[ii. Collision] \hfill \\
    The collision operator relaxes the populations toward equilibrium according to the relaxation equation:
    \begin{equation}
        f_i^{'} = f_i + \frac{1}{\tau}(f_i^{\mathrm{eq}} - f_i), 
    \end{equation}
which can be expressed in matrix form, $Ax = b$. Here, $A$ is a non-unitary diagonal matrix. To implement $A$ as a quantum operator, we use a two-term Linear Combination of Unitaries (LCU) decomposition \cite{Childs_Wiebe_2012}:
    \begin{equation}
        A = \frac{1}{2}(C_1 + C_2),
    \end{equation}
where $C_1 = A + i\sqrt{I - A^2}$ and $C_2 = A - i\sqrt{I - A^2}$ are unitary matrices. This decomposition requires a single ancilla qubit $|a\rangle$ and two Hadamard gates to act as the \textit{prepare} operators. The \textit{select} operation is implemented as a diagonal operator conditioned on the ancilla state, such that the state after collision is:
    \begin{equation}
        |\Psi_{D}\rangle = |0\rangle_a A |\Psi_{C}\rangle + |1\rangle_a \left(i\sqrt{I - A^2}\right) |\Psi_{C}\rangle.
    \end{equation}
    The physical result is obtained by post-selecting the ancilla in the $|0\rangle_a$ state.

    \item[iii. Streaming] \hfill \\
    In this step, we shift particle populations according to their velocities~\cite{Todorova_Steijl_2020},
    \begin{equation}
        f_i(\mathbf{x} + \mathbf{c}_i \Delta t, t + \Delta t) = f_i^{'}(\mathbf{x}, t),
    \end{equation}
which, on the quantum circuit, corresponds to controlled position shifts conditioned on the velocity register. The current implementation uses multiple Toffoli gates – a major algorithmic bottleneck in terms of scale.

    \item[iv. Measurement] \hfill \\
    The probability density function is reconstructed after repeated tomography measurements, and the zeroth moment is calculated to obtain the density and velocity fields as described in \cref{eq:moments}.
\end{description}

A summary of the typical algorithm is summarized in \cref{alg:qlbm}.

\begin{algorithm}
\caption{Typical Quantum Lattice Boltzmann Method (QLBM) Algorithm}
\label{alg:qlbm}
\begin{algorithmic}[1]
\STATE \textbf{Initialize} $|\psi\rangle_0$ with amplitude encoding of $\sqrt{f_i(x_j, 0)}$
\FOR{$t = 0$ \TO $N_{steps}-1$}
    \STATE \textbf{Collision:} Apply non-unitary relaxation $A$ via LCU using ancilla $|a\rangle$:
    \STATE \quad $|\psi\rangle_{coll} = \mathcal{U}_{coll} (|\psi\rangle_t \otimes |0\rangle_a)$
    \STATE \textbf{Streaming:} Perform controlled position shifts $\mathcal{S}_i$ conditioned on velocity:
    \STATE \quad $|\psi\rangle_{t+1} = \sum_{i} |i\rangle_v \langle i| \otimes \mathcal{S}_i |\psi\rangle_{coll}$
    \STATE \textbf{Measurement and Post-processing:} Perform full quantum state tomography and reconstruct the statevector to finally calculate $\rho(x, t+1)$ and $u(x, t+1)$ and visualize.
    \STATE \textbf{Update:} Reinitialize $\rho(x,t+1)$ as $\sqrt{f_i(x_j, t)}$ for the next step using new $u(x, t)$
\ENDFOR
\end{algorithmic}
\end{algorithm}

\subsection{Quantum Complexity Analysis}
In this paper, we discuss quantum computational complexity in terms of qubit number, transpiled circuit depth, and total transpiled two-qubit gate count. Transpilation is done at optimization level 3.

Let $M=2^{n_s}$ denote the number of lattice sites encoded using $n_s$ spatial qubits. The velocity register requires $n_v=\log_2b$ qubits, where $b=3$ from D1Q3. We also use one ancillary qubit to implement LCU during collision. The total qubit count for a simulation of $M$ lattices is $n = \log_2M + \log_2b + 1$. For larger domains, even with multiple dimensions and velocity directions, $M$ dominates. This confirms logarithmic qubit scaling.

Lastly, we note a major hurdle in implementing the general QLBM. Measurement of the final quantum state is done by full state tomography which scales as  $O(3^n)$. This has made the algorithm cumbersome to run and analyze on real hardware even for lattices as small as $M=32$.

\subsection{Implementation Details}
In this work, we implement a D1Q3 QLBM algorithm using Qiskit, with design choices motivated by near-term hardware constraints. Instead of implementing full state tomography with its exponential cost, we use the Qiskit \texttt{SamplerV2} primitive. Using the bitstring counts from total shots, we post-select on the $|0\rangle_a$ state of the ancilla qubit. We then renormalize the resulting amplitudes to obtain the macroscopic density fields. The algorithm is shown in \cref{alg:qlbm2}.

\begin{algorithm}
\caption{Quantum Lattice Boltzmann Method (QLBM) Timestep}
\label{alg:qlbm2}
\begin{algorithmic}[1]
\STATE \textbf{Initialize} $|\psi\rangle_0$ with amplitude encoding of $\sqrt{f_i(x_j, 0)}$
\FOR{$t = 0$ \TO $N_{steps}-1$}
    \STATE \textbf{Collision:} Apply non-unitary relaxation $A$ via LCU using ancilla $|a\rangle$:
    \STATE \quad $|\psi\rangle_{coll} = \mathcal{U}_{coll} (|\psi\rangle_t \otimes |0\rangle_a)$
    \STATE \textbf{Streaming:} Perform controlled position shifts $\mathcal{S}_i$ conditioned on velocity qubits:
    \STATE \quad $|\psi\rangle_{t+1} = \sum_{i} |i\rangle_v \langle i| \otimes \mathcal{S}_i |\psi\rangle_{coll}$
    \STATE \textbf{Measurement and Post-processing:} Execute \texttt{SamplerV2} and post-select on ancilla $|0\rangle_a$
    \STATE \textbf{Update:} Reinitialize $\rho(x,t+1)$ as $\sqrt{f_i(x_j, t)}$ for the next step, using new $u(x, t)$
\ENDFOR
\end{algorithmic}
\end{algorithm}

\section{Numerical Experiments}
\label{sec:experiments}

\subsection{Setup and configurations}
Setup begins with a normalized Gaussian initial condition. The target equation is the advection–diffusion model governed by a truncated BGK relaxation operator. All simulations employ periodic boundary conditions.  \Cref{tab:params} summarizes the parameters used throughout this study.

\begin{table}[htbp]
\footnotesize
\caption{Simulation parameters and lattice configurations.}
\label{tab:params}
\begin{center}
  \begin{tabular}{|c|c|c|c|} \hline
Parameter & Symbol & Value \\ \hline
Lattice size & $M$ & 4, 8, 16, 32, 64 \\
Advection velocity & $u$ & 0.1 \\
Spatial step & $\Delta x$ & 1 \\
Time step & $\Delta t$ & 1 \\
Shots & -- & 100000 \\
Spatial qubits & $n_s$ & $\log_2 M$ \\
Directional qubits & $n_v$ & 2 \\
Ancillary qubit & $n_a$ & 1 \\ \hline
\end{tabular}
\end{center}
\end{table}

\begin{figure}[H]
  \centering
  \includegraphics[width=1.0\textwidth]{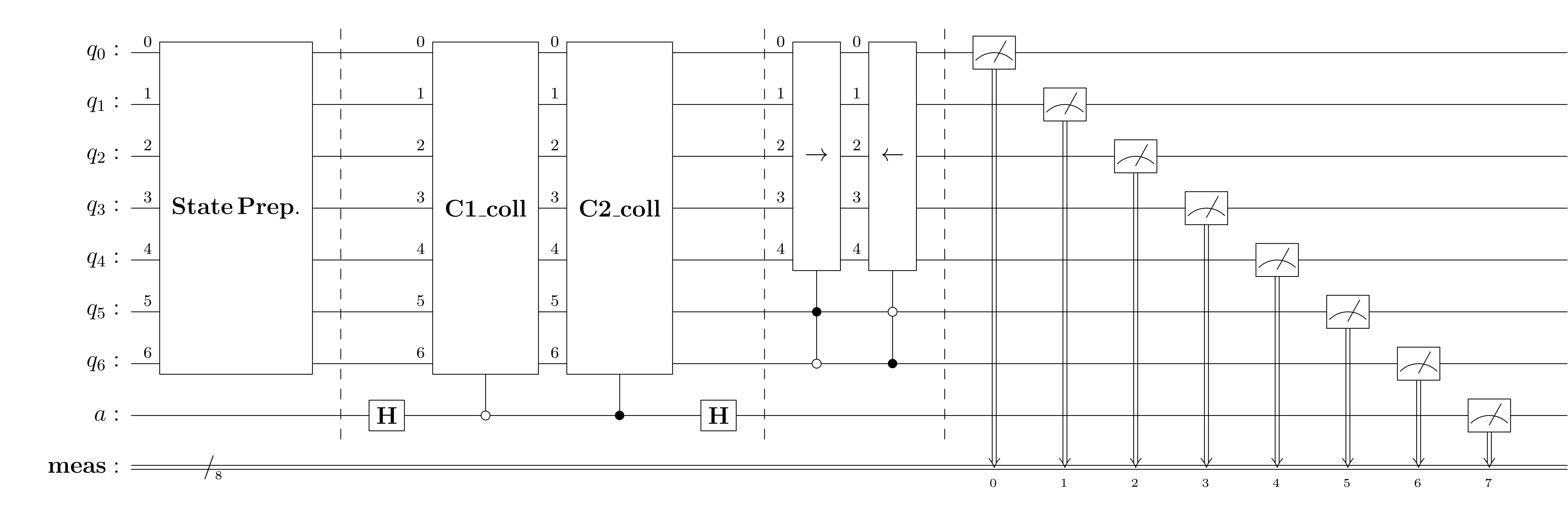}
  \caption{Sample QLBM circuit diagram for the M=32 case.}
  \label{fig:circuit32}
\end{figure}

In \cref{fig:circuit32}, we show a sample circuit for one timestep of QLBM on a $M=32$ grid. This yields a 8-qubit circuit showing the four main steps of the algorithm.

Each experiment proceeded through one timestep of the QLBM sequence - state preparation, collision, streaming, and measurement. Gate counts, circuit depth, and error calculations are recorded for each case.

\subsection{About the IBM Quantum Systems used}
The quantum simulations were executed across three distinct IBM Quantum systems to benchmark algorithmic performance across different hardware generations. We compared \textit{ibm\_rensselaer} (Quantum System One) against the newer System Two architectures, \textit{ibm\_kingston} and \textit{ibm\_boston}. It is important to note that the native gate sets of IBM Quantum System One and System Two are different. These are specified on the IBM Quantum cloud platform \cite{IBMQuantumPlatform2026}. The hardware specifications and calibration metrics at the time of execution are summarized in \cref{tab:hardware}. 
These metrics set the expectations for our hardware validation and dictate the limits of algorithmic scalability.
Two-Qubit Gate Error represents the general fidelity of entangling operations, such as ECR or CZ gates. Given that the QLBM requires many two-qubit operations \cite{Kocherla_Song_Chrit_Gard_Dumitrescu_Alexeev_Bryngelson_2024}, this parameter is the primary predictor of state dephasing. 
The Readout Error quantifies the probability of incorrect bitstring assignment during the measurement stage. Lower readout errors on the System Two (Heron) processors predict a more accurate reconstruction of the macroscopic density field during sampler-based post-processing.
Coherence Times ($T_1, T_2$), specifically energy relaxation ($T_1$) and phase coherence ($T_2$), define the timeframe during which quantum information remains viable. Circuits must complete execution well within these limits to avoid the loss of the encoded particle distribution functions. 
Finally, CLOPS (Circuit Layer Operations Per Second) characterizes the execution throughput of the QPU. The higher CLOPS ratings of the newer architectures directly result in the significantly reduced runtimes observed for larger lattice sizes. Based on this, we expect the Heron processors to have significantly shorter runtimes than the Eagle processor for the same cases.

\begin{table}[H]
\footnotesize
\caption{Comparative Quantum Hardware Specifications and Calibration Metrics}
\label{tab:hardware}
\begin{center}
  \begin{tabular}{|c|c|c|c|} \hline
Parameter & \textit{ibm\_rensselaer} & \textit{ibm\_kingston} & \textit{ibm\_boston} \\ \hline
System Generation & System One & System Two & System Two \\
Processor Type & Eagle (r3) & Heron (r2) & Heron (r3) \\
Qubit Count & 127 & 156 & 156 \\
Median 2Q Gate Error & $6.588 \times 10^{-3}$ & $1.841 \times 10^{-3}$ & $1.182 \times 10^{-3}$ \\
Median single-qubit (SX) Error & $2.41 \times 10^{-4}$ & $2.22 \times 10^{-4}$ & $1.60 \times 10^{-4}$ \\
Best 2Q Gate Error & $2.97 \times 10^{-2}$ (ECR) & $9.39 \times 10^{-4}$ (CZ) & $6.62 \times 10^{-4}$ (CZ) \\
Median Readout Error & $2.29 \times 10^{-2}$ & $9.52 \times 10^{-3}$ & $4.76 \times 10^{-3}$ \\
Median $T_1$ ($\mu$s) & 263.9 & 230.82 & 286.59 \\
Median $T_2$ ($\mu$s) & 174.4 & 126.45 & 327.64 \\
CLOPS & 260K & 340K & 340K \\ \hline
\end{tabular}
\end{center}
\end{table}

\subsection{Metrics of interest}
Performance and accuracy were quantified through both hardware metrics and physical error metrics derived from the simulation output.

\subsubsection*{Hardware and Algorithmic Quantities}
\begin{enumerate}
    \item Circuit Depth – The maximum number of gate layers in the transpiled circuit, representing the critical path length and a proxy for cumulative decoherence.
    \item Two-Qubit (2Q) Gate Count – The total number of entangling operations, which are the primary drivers of gate error in the solution.
    \item Quantum Processing Unit (QPU) Time – The wall-clock execution time measured for each experiment on the physical backend.
\end{enumerate}

\subsubsection*{Accuracy Measures}
\begin{enumerate}
    \item Mean Absolute Error ($\mathcal{L}_1$ error norm): This measures the average absolute deviation across all lattice sites.
    \begin{equation}
        \mathcal{L}_1 = \frac{1}{M}\|\rho_{q} - \rho_{\mathrm{sv}}\|_1 = \frac{1}{M} \sum_{j=1}^{M} |\rho_{q,j} - \rho_{\mathrm{sv},j}|.
    \end{equation}
    
    \item Maximum Absolute Error ($\mathcal{L}_\infty$ error norm): This is used to identify the worst-case local deviations between the computed density $\rho_{q}$ and the ideal statevector density $\rho_{\mathrm{sv}}$.
    \begin{equation}
        \mathcal{L}_\infty = \|\rho_{q} - \rho_{\mathrm{sv}}\|_\infty = \max_{1\leq j \leq M} |\rho_{q,j} - \rho_{\mathrm{sv},j}|.
    \end{equation}
    
    \item Root-Mean-Square Error (related to $\mathcal{L}_2$ error norm): This serves as a global measure of the hardware result noise, indicating the total drift of the computed solution from the ideal.
    \begin{equation}
        \mathrm{RMSE} = \frac{1}{\sqrt{M}}\|\rho_{q} - \rho_{\mathrm{sv}}\|_2 = \sqrt{\frac{1}{M} \sum_{j=1}^{M} (\rho_{q,j} - \rho_{\mathrm{sv},j})^2}.
    \end{equation}
    
\end{enumerate}

The three accuracy metrics are reported to provide a comprehensive view of the error distribution: the $\mathcal{L}_1$ error represents the average global accuracy of the density field, the $\mathcal{L}_\infty$ error identifies the worst-case local deviations, and the RMSE emphasizes larger outliers caused by the noise. 

Together, they characterize both resource scaling and solution accuracy as lattice size increases, enabling a quantitative assessment of QLBM performance on present-day NISQ hardware.

\section{Results and Discussions}
\label{sec:results}

\subsection{Algorithm Complexity Results}
We analyze the algorithm in terms of circuit depth at optimization level 3 according to the native gate set of the IBM Heron processor: $\{cz, id, rx, rz, rzz, sx, x\}$. The observed total circuit depth and two-qubit gate count, shown in \cref{tab:complexity}, grow approximately as $O(M^2)$. Importantly, no exponential depth growth is observed, confirming that circuit construction remains scalable and that performance limitations can be improved with hardware advancements.

\begin{table}[H]
\footnotesize
    \caption{Transpiled circuit complexity across lattice sizes ($M$) at optimization level 3.}
    \label{tab:complexity}
\begin{center} 
    \begin{tabular}{|c|c|c|c|}
    \hline
    \textbf{Lattice Size ($M$)} & \textbf{Qubits} & \textbf{Total Circuit Depth} & \textbf{Two-qubit Gate Count} \\ \hline
    4  & 5 & 152   & 61    \\ 
    8  & 6 & 927   & 402   \\ 
    16 & 7 & 3,175  & 1,329  \\ 
    32 & 8 & 10,508 & 4,404  \\ 
    64 & 9 & 34,300 & 14,316 \\ \hline
\end{tabular}
\end{center}
\end{table}

\begin{figure}[H]
  \centering
  \includegraphics[width=0.8\textwidth]{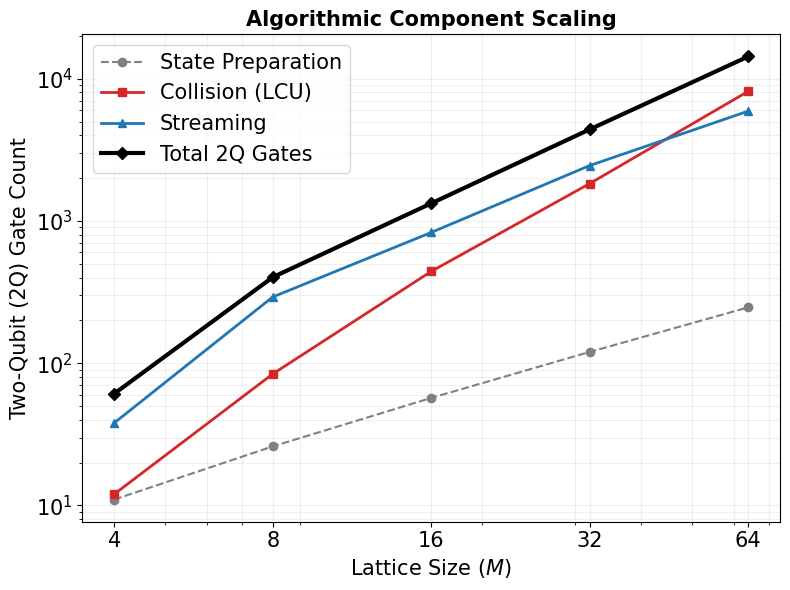}
  \caption{Scaling of two-qubit (2Q) gate counts across lattice sizes $M \in \{4, 8, 16, 32, 64\}$. }
  \label{fig:components}
\end{figure}

The plot in \cref{fig:components} shows the individual contributions of the algorithm steps.
The log--log representation highlights a transition in algorithmic dominance: while streaming (triangles) is the primary resource consumer at lower resolutions, the collision operator (squares) exhibits super-linear growth and becomes the dominant bottleneck at $M=64$. The steep progression in the total 2Q gate count (diamonds) identifies the collision step as the critical factor in noise accumulation for large-scale QLBM simulations.

\subsection{Verification and Density Profiles}
\begin{figure}[H]
  \centering
  \includegraphics[width=0.9\textwidth]{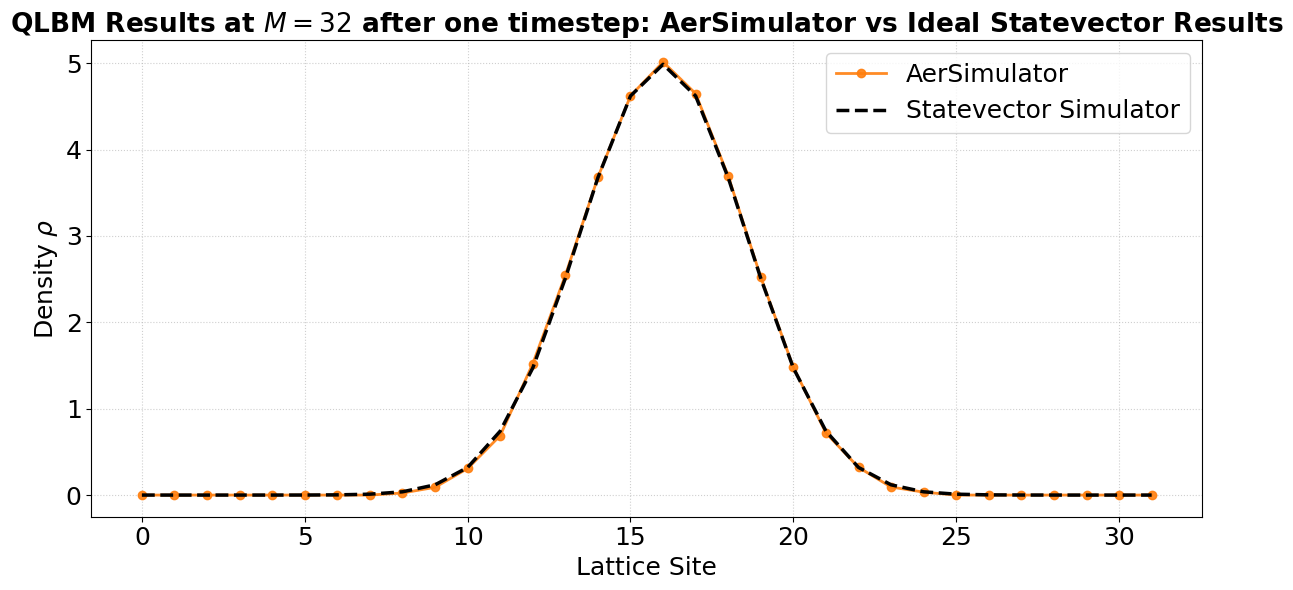}
  \caption{Comparison of reconstructed macroscopic density from the AerSimulator (sampled) against the ideal Statevector solution for $M=32$ after one timestep.}
  \label{fig:simsim_32}
\end{figure}

To assess solution accuracy, the macroscopic density field reconstructed from our sampled result was first compared to the reference statevector simulator solution after one timestep for each lattice size $M$. \Cref{fig:simsim_32} shows a representative case for $M=32$, where the result obtained through the Qiskit AerSimulator backend is plotted against the Qiskit Statevector Simulator. The close agreement between the plots validates the accuracy of our sampler-based algorithm, thereby enabling hardware validation. 

\begin{figure}[H]
  \centering
  \includegraphics[width=0.9\textwidth]{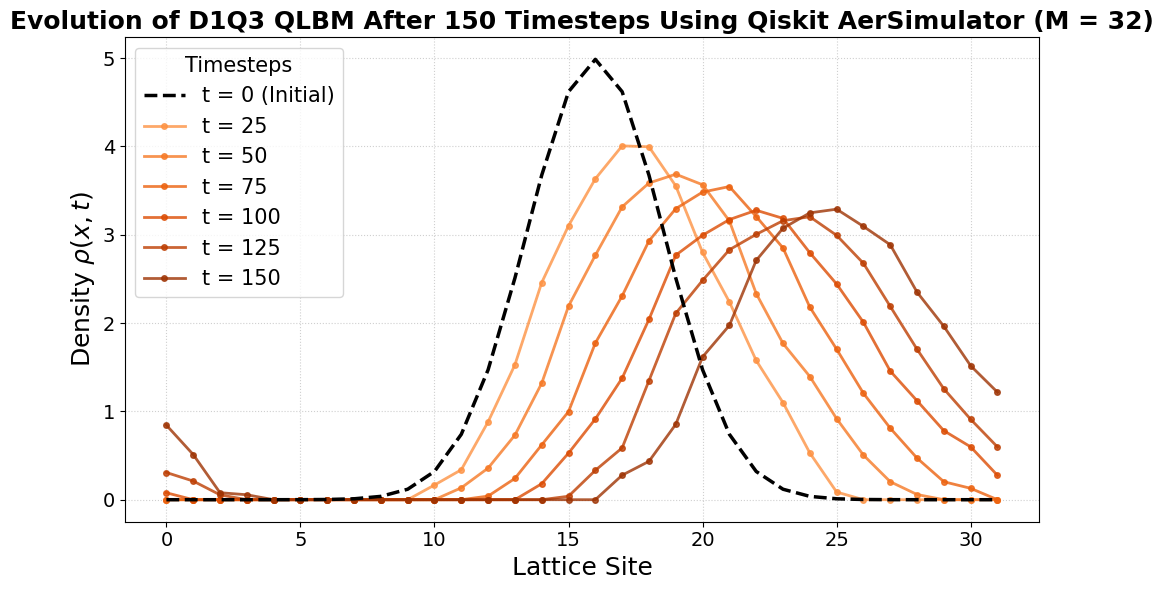}
  \caption{Evolution of the macroscopic density field after 150 timesteps, illustrating the sampler-based advection and diffusion of the initial Gaussian profile at 100,000 shots using the AerSimulator backend.}
  \label{fig:multistep}
\end{figure}

The temporal evolution of the macroscopic density field is captured in \cref{fig:multistep}. After 150 timesteps, the initial Gaussian profile qualitatively exhibits both advection in the direction of the prescribed velocity $u$ and physical spreading due to diffusion as governed by the underlying BGK relaxation dynamics. The quantitative variations between \cref{fig:multistep} and the reference solution in \cref{fig:gaussian_hill} result from sampling errors accumulated over the timesteps.

In this study, we restrict our hardware analysis to a single timestep to isolate the fundamental scaling characteristics of the QLBM operators. A primary constraint is the accumulation of incoherent noise; since circuit depth grows linearly with the number of timesteps, the $O(10^4)$ gate depth already observed at $M=64$ in \cref{fig:components} for a single step would quickly exceed the $T_2$ coherence limits of NISQ devices in a multi-step simulation. Furthermore, the limits of the sampler-based measurement strategy impose significant challenges for temporal scaling. Because the LCU implementation is inherently non-unitary, the resulting probability distribution requires specific post-processing, including rescaling and renormalization to reconstruct physical density fields. In a multi-step execution, the requirement to re-encode these hardware-distorted and rescaled results as the initial state for the subsequent step would necessitate an interpolation of the amplitudes, potentially introducing significant numerical errors that compound over time.

\begin{figure}[H]
  \centering
  \begin{minipage}{0.48\textwidth}
    \centering
    \includegraphics[width=\textwidth]{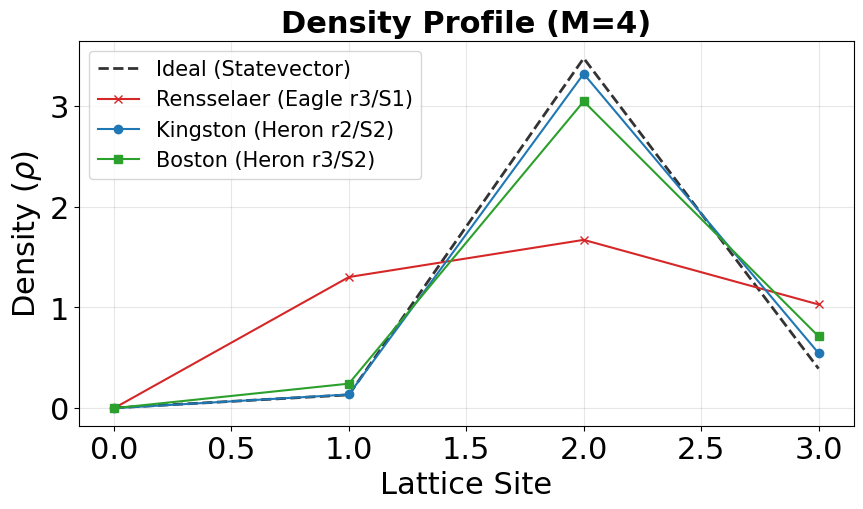}
    \\ (a) $M=4$
  \end{minipage}
  \hfill
  \begin{minipage}{0.48\textwidth}
    \centering
    \includegraphics[width=\textwidth]{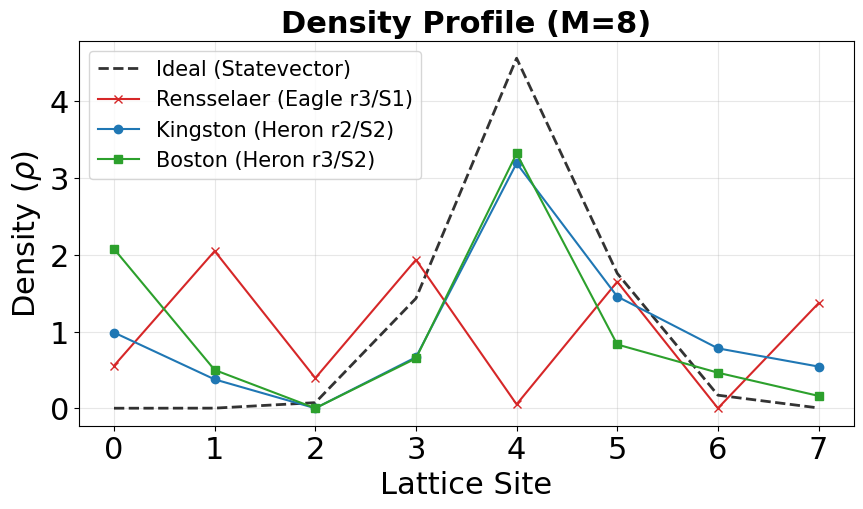}
    \\ (b) $M=8$
  \end{minipage}
  \vspace{0.3cm}
  \begin{minipage}{0.48\textwidth}
    \centering
    \includegraphics[width=\textwidth]{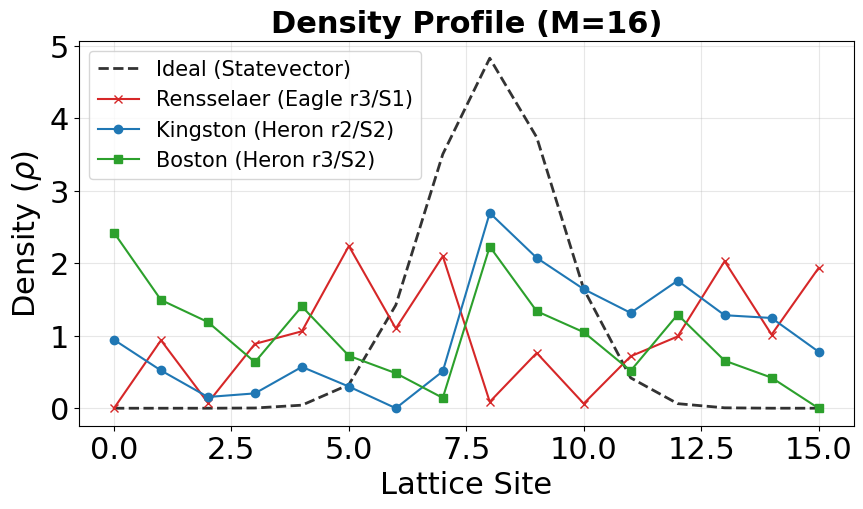}
    \\ (c) $M=16$
  \end{minipage}
  \hfill
  \begin{minipage}{0.48\textwidth}
    \centering
    \includegraphics[width=\textwidth]{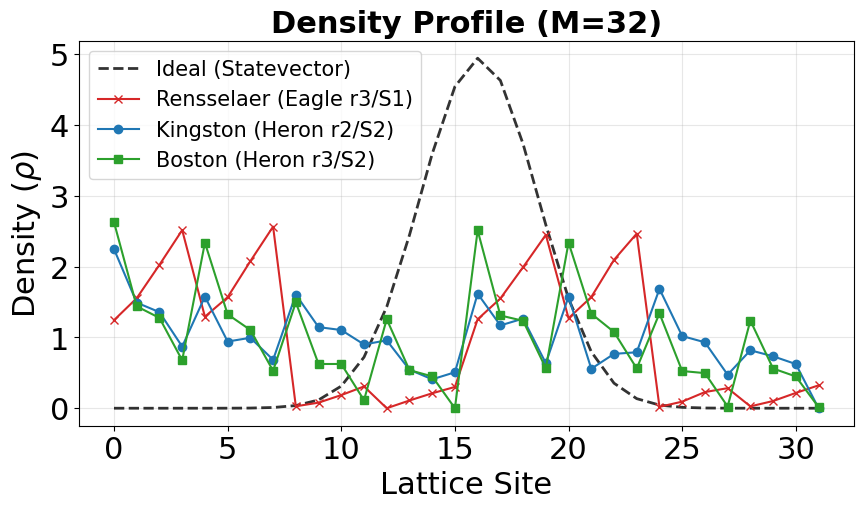}
    \\ (d) $M=32$
  \end{minipage}
  \vspace{0.3cm}
  \begin{minipage}{0.48\textwidth}
    \centering
    \includegraphics[width=\textwidth]{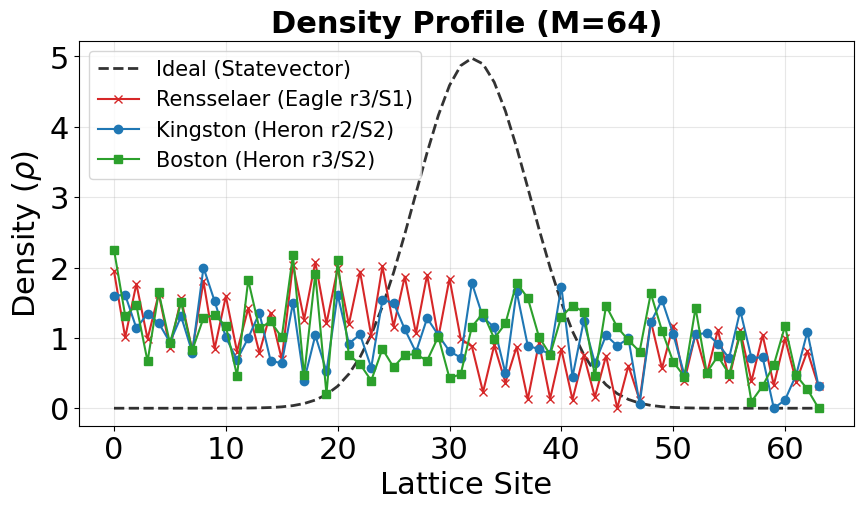}
    \\ (e) $M=64$
  \end{minipage}
  \caption{Accuracy comparison between the Qiskit Statevector simulator (dashed line) and IBM Quantum System Two hardware (solid line) across various lattice sizes $M$. The hardware reproduces the advection--diffusion profile, with deviations arising from decoherence and gate noise.}
  \label{fig:density_all_M}
\end{figure}

The hardware results shown in \cref{fig:density_all_M} qualitatively reproduce the advection--diffusion profile. However, deviations are visible near the domain boundaries, where decoherence distorts the amplitudes. These discrepancies grow with lattice size, $M$. This observation is consistent with \cref{tab:complexity} as the circuit depth and the two-qubit gate count increase significantly with $M$, leading to increased cumulative readout noise and two-qubit gate errors.

For each $M$, observations are made on the density plots obtained from the different hardware backends. As predicted by the calibration data in \cref{tab:hardware}, the newer \textit{ibm\_kingston} and \textit{ibm\_boston} backends perform significantly better than the \textit{ibm\_rensselaer} backend. This is a combined result of superior qubit decoherence times, lower readout errors, and optimized two-qubit gate counts that arise from different sets of native gates. Specifically, enhanced coherence times ($T_1, T_2$) reduce the rate of phase and energy relaxation during the execution of deep QLBM circuits, thereby preserving the structural integrity of the encoded distribution functions. Lower readout error rates minimize the systematic shifts in probability amplitudes during the measurement stage, ensuring a more accurate reconstruction of the macroscopic density field from the sampled counts. Furthermore, the optimized native gate sets of the Heron-based processors allow for more efficient decomposition of the QLBM operators, reducing the total number of two-qubit operations and the associated accumulation of noise induced by multiple controlled-not (CNOT) gates.

The preservation of the total macroscopic density is a fundamental requirement for the physical validity of the simulation. Theoretically, the update rule is
designed to be mass-conserving, meaning the sum of the distribution functions should recover to the initial total mass. 
In our sampler-based implementation, we observe that for up to $M \le 16$, the total density is qualitatively preserved across the domain for the Heron processor output. However, as the lattice size $M$ increases to $32$ and $64$, the high two-qubit (2Q) gate count, as shown in \cref{fig:components}, introduces significant noise that causes a ``leakage'' of information into the non-physical or ``garbage'' subspaces of the full statevector. This results in a reconstructed total density that is lower than the theoretical unity, as the sampler primarily captures bitstrings from the successful post-selection on the ancilla qubit $|0\rangle_a$, which now carries a reduced total amplitude due to decoherence.

Despite these variations, the underlying physics is largely preserved across hardware runs, validating that the implemented collision, streaming, and postprocessing subroutines correctly encode the QLBM update rules.

\subsection{Quantitative Scaling Analysis}
\begin{figure}[H]
  \centering
  \begin{minipage}{0.48\textwidth}
    \centering
    \includegraphics[width=\textwidth]{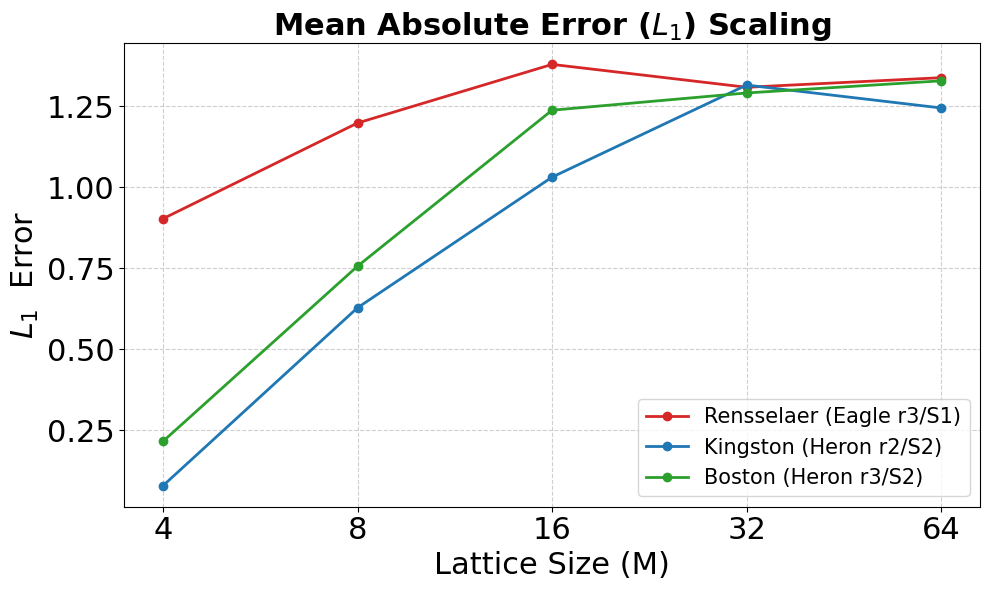}
    \\ (a)
    \label{fig:l1}
  \end{minipage}
  \hfill
  \begin{minipage}{0.48\textwidth}
    \centering
    \includegraphics[width=\textwidth]{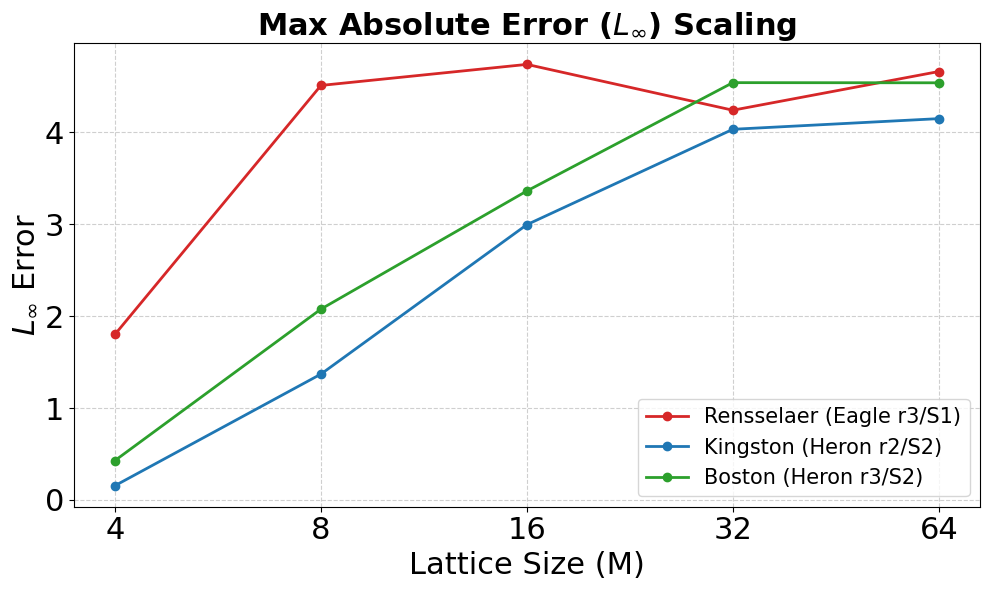}
    \\ (b)
    \label{fig:mae}
  \end{minipage}

  \vspace{0.5cm} 

  \begin{minipage}{0.48\textwidth}
    \centering
    \includegraphics[width=\textwidth]{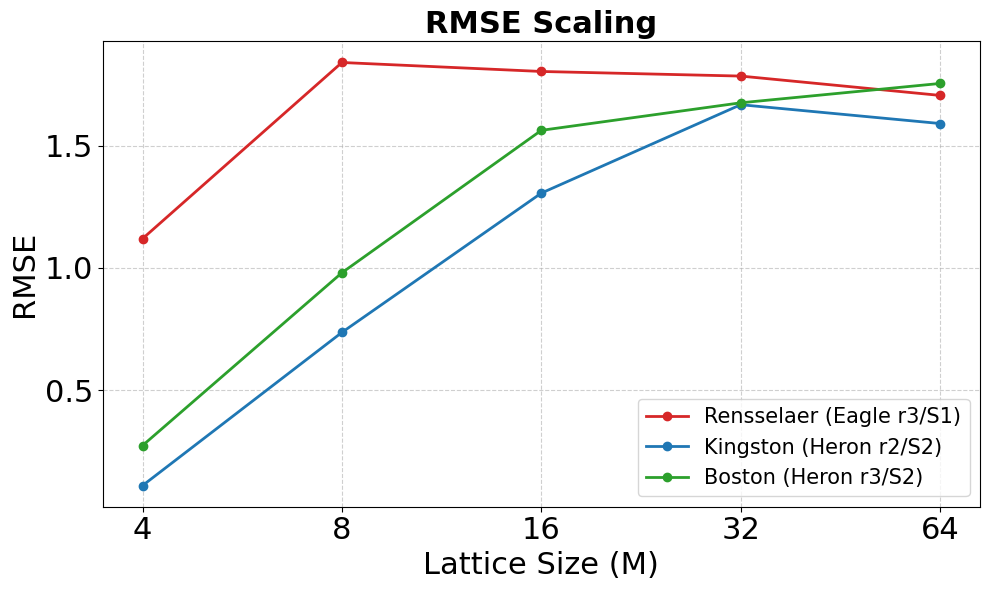}
    \\ (c)
    \label{fig:rmse}
  \end{minipage}

  \caption{Quantitative performance and error scaling of the QLBM algorithm across lattice sizes $M \in \{4, 8, 16, 32, 64\}$ for one timestep. These subfigures illustrate the scaling of physical error metrics between the three hardware results, showing a characteristic increase in error as circuit depth grows with $M$.}
  \label{fig:summary}
  \end{figure}

A quantitative summary of the algorithmic performance is presented in \cref{fig:summary}. The plots show the scaling of the $\mathcal{L}_1$ error, $\mathcal{L}_\infty$ error, and RMSE index across the three quantum backends. 

In all subplots, we observe a clear performance hierarchy for lower $M$: the System Two architectures (\textit{ibm\_kingston} and \textit{ibm\_boston}) consistently outperform the System One (\textit{ibm\_rensselaer}) backend. The lower median two-qubit gate errors of the Heron processors in \textit{ibm\_boston} and \textit{ibm\_kingston} result in significantly lower errors compared to the Eagle r3 processor.

As $M$ increases beyond $32$, we observe distinct crossovers in the error scaling plots. This behavior is consistent with the density profiles presented in \cref{fig:density_all_M}, where hardware noise begins to dominate the reconstructed amplitudes, effectively damping any meaningful physical correlations. This confirms that for the current NISQ generation, $M=32$ represents a practical upper bound for resolving 1D advection--diffusion features on the Heron and Eagle architectures. Still, the overall correlation indicates that improvements in hardware fidelity directly translate into higher accuracy in QCFD simulations, up to the point where circuit-depth-induced decoherence becomes the limiting factor.

\begin{figure}[htbp]
  \centering
  \includegraphics[width=0.8\textwidth]{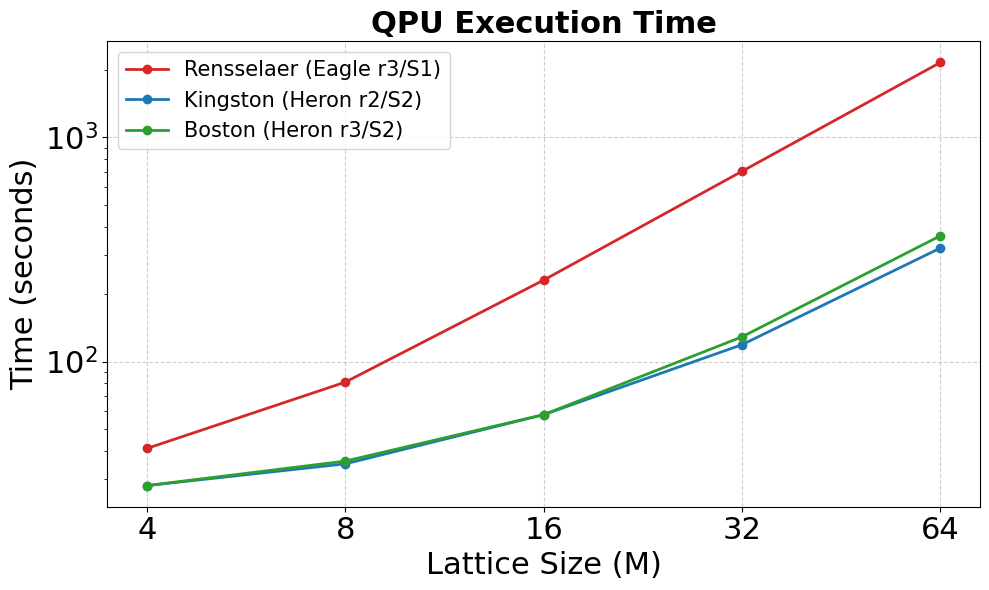}
  \caption{Comparison of trend in total QPU time for one timestep of QLBM with increasing $M$ across the different quantum hardware (shots = 100,000).}
  \label{fig:qpu}
\end{figure}

Finally, \cref{fig:qpu} reports the scaling of QPU runtime. In this log--log plot, all three backends display a similar positive power-law scaling, consistent with the logarithmic increase in qubit requirements and polynomial gate depth scaling of the QLBM algorithm. For an equal number of shots, the \textit{ibm\_rensselaer} baseline exhibits the highest absolute runtime, while the System Two results (\textit{ibm\_kingston} and \textit{ibm\_boston}) are distinguished by a significant downward vertical shift. 
The relationship between the higher CLOPS rating of the newer generation System Two machines and the reduced QPU time requirement underscores the improved computational throughput of newer hardware generations. These performance gains suggest that as processor efficiency scales, high-fidelity QLBM simulations will become increasingly feasible for practical, near-term research applications.

\Cref{tab:comprehensive_benchmarks} presents a summary of all the performance metrics discussed in this section.

\begin{table}[htbp]
\centering
\caption{Comprehensive Benchmark of QLBM Performance. Data compares resource requirements (Qubits), execution efficiency (QPU Time), and physical accuracy (RMSE, $\mathcal{L}_\infty$ error, $\mathcal{L}_1$ error) across three generations of IBM Quantum hardware.}
\label{tab:comprehensive_benchmarks}
\begin{small}
\begin{tabular}{|l|c|l|r|c|c|c|}
\hline
$M$ & Qubits & Backend & QPU Time (s) & RMSE & $\mathcal{L}_\infty$ error & $\mathcal{L}_1$ error \\ \hline
4  & 5 & \textit{ibm\_rensselaer} & 41   & 1.1199 & 1.8016 & 0.9026 \\
   &   & \textit{ibm\_kingston}    & 28   & 0.1092 & 0.1545 & 0.0790 \\
   &   & \textit{ibm\_boston}      & 28   & 0.2731 & 0.4281 & 0.2158 \\ \hline
8  & 6 & \textit{ibm\_rensselaer} & 81   & 1.8412 & 4.5120 & 1.1977 \\
   &   & \textit{ibm\_kingston}    & 35   & 0.7363 & 1.3691 & 0.6279 \\
   &   & \textit{ibm\_boston}      & 36   & 0.9804 & 2.0766 & 0.7561 \\ \hline
16 & 7 & \textit{ibm\_rensselaer} & 231  & 1.8045 & 4.7411 & 1.3785 \\
   &   & \textit{ibm\_kingston}    & 58   & 1.3065 & 2.9962 & 1.0307 \\
   &   & \textit{ibm\_boston}      & 58   & 1.5634 & 3.3633 & 1.2372 \\ \hline
32 & 8 & \textit{ibm\_rensselaer} & 706  & 1.7856 & 4.2422 & 1.3080 \\
   &   & \textit{ibm\_kingston}    & 119  & 1.6684 & 4.0346 & 1.3148 \\
   &   & \textit{ibm\_boston}      & 129  & 1.6762 & 4.5425 & 1.2904 \\ \hline
64 & 9 & \textit{ibm\_rensselaer} & 2,159 & 1.7065 & 4.6635 & 1.3375 \\
   &   & \textit{ibm\_kingston}    & 321  & 1.5913 & 4.1506 & 1.2441 \\
   &   & \textit{ibm\_boston}      & 364  & 1.7554 & 4.5408 & 1.3278 \\ \hline
\end{tabular}
\end{small}
\end{table}

\section{Conclusion and Future Directions}
\label{sec:conclusions}

In this work, we successfully implemented and benchmarked a one-dimensional QLBM algorithm for the advection--diffusion equation on three generations of IBM Quantum hardware. By employing a sampler-based measurement strategy and the LCU approach for the collision operator, we demonstrated that the algorithm effectively reproduced the macroscopic physics of transport and spreading processes within reasonable limits of NISQ computers. Our results confirm a clear performance hierarchy: the newer System Two architectures, \textit{ibm\_kingston} and \textit{ibm\_boston}, provide significantly higher fidelity than the System One \textit{ibm\_rensselaer} backend, with errors directly correlating to calibration data.

Despite these promising results, several pathways for future research remain. While the current study focused on a single timestep to isolate hardware noise effects, future work will explore the cumulative impact of these errors over longer temporal evolutions. As quantum hardware approaches the fault-tolerant era, we anticipate that reduced gate errors, higher CLOPS ratings, and increased coherence times will allow for the resolution of more complex flow features without the rapid loss of state fidelity observed in the $M=64$ cases.

Furthermore, the modular nature of the QLBM framework allows for several algorithmic extensions. First, the introduction of non-linear collision operators via Carleman linearization~\cite{Itani_Sreenivasan_Succi_2024} could enable the simulation of the full BGK operator, extending the method to high-Reynolds-number flows. Second, the mapping of the D2Q9 or D3Q27 lattice velocity sets would allow for the resolution of two- and three-dimensional fluid domains, though this will require a commensurate increase in qubit counts and connectivity. Finally, benchmarking this QLBM formulation against other quantum modalities, such as trapped-ion or neutral-atom processors, could provide valuable insights into the optimal hardware architecture for fluid dynamics.

\section*{Acknowledgments}
The authors would like to acknowledge the use of IBM Quantum services for this work. Quantum jobs were executed on the \textit{ibm\_rensselaer}, \textit{ibm\_kingston}, and \textit{ibm\_boston} backends. We thank IBM Quantum for their support under the collaboration executed between Rensselaer Polytechnic Institute and IBM under the terms of the Artificial Intelligence Research Collaboration Agreement Number W1771793.

 The authors used Anthropic Claude to assist with editing the manuscript for readability and with refactoring, documenting, and testing the accompanying code. The authors assume responsibility for all content.

\bibliographystyle{siamplain}
\bibliography{references}

\begin{thebibliography}{10}

\bibitem{Budinski_2021}
{\sc L.~Budinski}, {\em Quantum algorithm for the advection–diffusion equation simulated with the lattice boltzmann method}, Quantum Information Processing, 20 (2021), p.~57, \url{https://doi.org/10.1007/s11128-021-02996-3}, \url{http://link.springer.com/10.1007/s11128-021-02996-3}.

\bibitem{Childs_Wiebe_2012}
{\sc A.~M. Childs and N.~Wiebe}, {\em Hamiltonian simulation using linear combinations of unitary operations},  (2012), \url{https://doi.org/10.48550/arXiv.1202.5822}, \url{http://arxiv.org/abs/1202.5822}.
\newblock arXiv:1202.5822.

\bibitem{Harrow_Hassidim_Lloyd_2009}
{\sc A.~W. Harrow, A.~Hassidim, and S.~Lloyd}, {\em Quantum algorithm for linear systems of equations}, Physical Review Letters, 103 (2009), p.~150502, \url{https://doi.org/10.1103/PhysRevLett.103.150502}, \url{https://link.aps.org/doi/10.1103/PhysRevLett.103.150502}.

\bibitem{IBMQuantumPlatform2026}
{\sc {IBM Quantum}}, {\em Compute resources | {IBM} {Quantum} {Platform}}, 2026, \url{https://quantum.cloud.ibm.com/computers}.
\newblock Accessed: 2026-02-18.

\bibitem{Itani_Sreenivasan_Succi_2024}
{\sc W.~Itani, K.~R. Sreenivasan, and S.~Succi}, {\em Quantum algorithm for lattice boltzmann (qalb) simulation of incompressible fluids with a nonlinear collision term}, Physics of Fluids, 36 (2024), p.~017112, \url{https://doi.org/10.1063/5.0176569}, \url{https://pubs.aip.org/pof/article/36/1/017112/2933116/Quantum-algorithm-for-lattice-Boltzmann-QALB}.

\bibitem{Kocherla_Song_Chrit_Gard_Dumitrescu_Alexeev_Bryngelson_2024}
{\sc S.~Kocherla, Z.~Song, F.~E. Chrit, B.~Gard, E.~F. Dumitrescu, A.~Alexeev, and S.~H. Bryngelson}, {\em Fully quantum algorithm for mesoscale fluid simulations with application to partial differential equations}, AVS Quantum Science, 6 (2024), p.~033806, \url{https://doi.org/10.1116/5.0217675}, \url{https://pubs.aip.org/aqs/article/6/3/033806/3313544/Fully-quantum-algorithm-for-mesoscale-fluid}.

\bibitem{Kolmogorov_1991}
{\sc A.~N. Kolmogorov}, {\em The local structure of turbulence in incompressible viscous fluid for very large reynolds numbers}, Proceedings of the Royal Society of London. Series A: Mathematical and Physical Sciences, 434 (1991), p.~9–13, \url{https://doi.org/10.1098/rspa.1991.0075}, \url{https://royalsocietypublishing.org/rspa/article/434/1890/9/16588/The-local-structure-of-turbulence-in}.

\bibitem{Kruger_Kusumaatmaja_Kuzmin_Shardt_Silva_Viggen_2017}
{\sc T.~Kr\"uger, H.~Kusumaatmaja, A.~Kuzmin, O.~Shardt, G.~Silva, and E.~M. Viggen}, {\em The Lattice Boltzmann Method: Principles and Practice}, Graduate Texts in Physics, Springer International Publishing, Cham, 2017, \url{https://doi.org/10.1007/978-3-319-44649-3}, \url{http://link.springer.com/10.1007/978-3-319-44649-3}.

\bibitem{Kyriienko_Paine_Elfving_2021}
{\sc O.~Kyriienko, A.~E. Paine, and V.~E. Elfving}, {\em Solving nonlinear differential equations with differentiable quantum circuits}, Physical Review A, 103 (2021), p.~052416, \url{https://doi.org/10.1103/PhysRevA.103.052416}, \url{https://link.aps.org/doi/10.1103/PhysRevA.103.052416}.

\bibitem{Liu_Kolden_Krovi_Loureiro_Trivisa_Childs_2021}
{\sc J.-P. Liu, H.~{\O}. Kolden, H.~K. Krovi, N.~F. Loureiro, K.~Trivisa, and A.~M. Childs}, {\em Efficient quantum algorithm for dissipative nonlinear differential equations}, Proceedings of the National Academy of Sciences, 118 (2021), p.~e2026805118, \url{https://doi.org/10.1073/pnas.2026805118}, \url{https://pnas.org/doi/full/10.1073/pnas.2026805118}.

\bibitem{Lubasch_Joo_Moinier_Kiffner_Jaksch_2020}
{\sc M.~Lubasch, J.~Joo, P.~Moinier, M.~Kiffner, and D.~Jaksch}, {\em Variational quantum algorithms for nonlinear problems}, Physical Review A, 101 (2020), p.~010301, \url{https://doi.org/10.1103/PhysRevA.101.010301}, \url{https://link.aps.org/doi/10.1103/PhysRevA.101.010301}.

\bibitem{Lacatus_Moller_2025}
{\sc M.~Lăcătuş and M.~Möller}, {\em Surrogate quantum circuit design for the lattice boltzmann collision operator},  (2025), \url{https://doi.org/10.48550/ARXIV.2507.12256}, \url{https://arxiv.org/abs/2507.12256}.

\bibitem{Malinverno_2023}
{\sc G.~Malinverno}, {\em Quantum computing cfd simulations: state of the art}, Dec. 2023, p.~174–178, \url{https://doi.org/10.21741/9781644902813-38}, \url{https://www.mrforum.com/product/9781644902813-38}.

\bibitem{Mohamad_2019}
{\sc A.~A. Mohamad}, {\em Lattice Boltzmann Method: Fundamentals and Engineering Applications with Computer Codes}, Springer London, London, 2019, \url{https://doi.org/10.1007/978-1-4471-7423-3}, \url{http://link.springer.com/10.1007/978-1-4471-7423-3}.

\bibitem{Moin_Mahesh_1998}
{\sc P.~Moin and K.~Mahesh}, {\em Direct numerical simulation: A tool in turbulence research}, Annual Review of Fluid Mechanics, 30 (1998), p.~539–578, \url{https://doi.org/10.1146/annurev.fluid.30.1.539}, \url{https://www.annualreviews.org/doi/10.1146/annurev.fluid.30.1.539}.

\bibitem{Nielsen_Chuang_2010}
{\sc M.~A. Nielsen and I.~L. Chuang}, {\em Quantum computation and quantum information}, Cambridge university press, Cambridge, 10th anniversary edition~ed., 2010.

\bibitem{Pope_2015}
{\sc S.~B. Pope}, {\em Turbulent flows}, Cambridge Univ. Press, Cambridge, 1. publ., 12. print~ed., 2015.

\bibitem{Shinde_Budinski_Niemimaki_Lahtinen_Liebelt_Li_2025}
{\sc T.~Shinde, L.~Budinski, O.~Niemim\"aki, V.~Lahtinen, H.~Liebelt, and R.~Li}, {\em Utilizing classical programming principles in the intel quantum sdk: implementation of quantum lattice boltzmann method}, ACM Transactions on Quantum Computing, 6 (2025), p.~1–18, \url{https://doi.org/10.1145/3678185}, \url{https://dl.acm.org/doi/10.1145/3678185}.

\bibitem{Spalart_2000}
{\sc P.~Spalart}, {\em Strategies for turbulence modelling and simulations}, International Journal of Heat and Fluid Flow, 21 (2000), p.~252–263, \url{https://doi.org/10.1016/S0142-727X(00)00007-2}, \url{https://linkinghub.elsevier.com/retrieve/pii/S0142727X00000072}.

\bibitem{Syamlal_Copen_Takahashi_Hall_2024}
{\sc M.~Syamlal, C.~Copen, M.~Takahashi, and B.~Hall}, {\em Computational fluid dynamics on quantum computers}, in AIAA AVIATION FORUM AND ASCEND 2024, Las Vegas, Nevada, 2024, American Institute of Aeronautics and Astronautics, \url{https://doi.org/10.2514/6.2024-3534}, \url{https://arc.aiaa.org/doi/10.2514/6.2024-3534}.

\bibitem{Tiwari_Iaconis_Jojo_Ray_Roetteler_Hill_Pathak_2025}
{\sc A.~Tiwari, J.~Iaconis, J.~Jojo, S.~Ray, M.~Roetteler, C.~Hill, and J.~Pathak}, {\em Algorithmic advances towards a realizable quantum lattice boltzmann method},  (2025), \url{https://doi.org/10.48550/arXiv.2504.10870}, \url{http://arxiv.org/abs/2504.10870}.
\newblock arXiv:2504.10870.

\bibitem{Todorova_Steijl_2020}
{\sc B.~N. Todorova and R.~Steijl}, {\em Quantum algorithm for the collisionless boltzmann equation}, Journal of Computational Physics, 409 (2020), p.~109347, \url{https://doi.org/10.1016/j.jcp.2020.109347}, \url{https://linkinghub.elsevier.com/retrieve/pii/S0021999120301212}.

\bibitem{Wang_Meng_Zhao_Yang_2025}
{\sc B.~Wang, Z.~Meng, Y.~Zhao, and Y.~Yang}, {\em Quantum lattice boltzmann method for simulating nonlinear fluid dynamics}, npj Quantum Information, 11 (2025), p.~196, \url{https://doi.org/10.1038/s41534-025-01142-6}, \url{https://www.nature.com/articles/s41534-025-01142-6}.

\bibitem{Wawrzyniak_Winter_Schmidt_Indinger_Janssen_Schramm_Adams_2025}
{\sc D.~Wawrzyniak, J.~Winter, S.~Schmidt, T.~Indinger, C.~F. Janßen, U.~Schramm, and N.~A. Adams}, {\em A quantum algorithm for the lattice-boltzmann method advection-diffusion equation}, Computer Physics Communications, 306 (2025), p.~109373, \url{https://doi.org/10.1016/j.cpc.2024.109373}, \url{https://linkinghub.elsevier.com/retrieve/pii/S0010465524002960}.

\bibitem{Yepez_2001}
{\sc J.~Yepez}, {\em Quantum lattice-gas model for computational fluid dynamics}, Physical Review E, 63 (2001), p.~046702, \url{https://doi.org/10.1103/PhysRevE.63.046702}, \url{https://link.aps.org/doi/10.1103/PhysRevE.63.046702}.

\end{thebibliography}
\end{document}